# A NEW BEAM INJECTION SCHEME FOR THE FERMILAB BOOSTER *

C. M. Bhat[#]
Fermilab, Batavia, IL 60510, USA

*Abstract*
A new beam injection scheme is proposed for the Fermilab Booster to increase beam brightness. The beam is injected on the deceleration part of the sinusoidal magnetic ramp and capture is started immediately after the injection. During the entire capture process we impose $\dot{P} = 0$ in a changing $B$ field. Beam dynamics simulations clearly show that this method is very efficient with no longitudinal beam emittance dilution and no beam loss. As a consequence of preserved emittance, the required RF power on a typical Booster cycle can be reduced by ~30% as compared with the scheme in current operation. Further, we also propose snap bunch rotation at extraction to reduce $dP/P$ of the beam to improve the slip-stacking efficiency in MI/RR.

## INTRODUCTION

The Fermilab Booster is one of the oldest rapid cycling proton synchrotron [1, 2] in the world. It uses a 15 Hz sinusoidal magnetic ramp for beam acceleration. The Booster receives $H^-$ beam of 400 MeV kinetic energy from the LINAC while it is at $B_{min}$ (minimum of the magnetic field ramp). $H^-$ charge exchange injection scheme is adopted to accumulate multi-turn proton beam in the ring. Depending on the number of Booster Turns (BT), the duration of the beam injection varies in the range of 2-40 μsec. Ever since the Booster came into operation, the beam injection was carried out in the region of fairly constant magnetic field of the ramp i.e., close to $B_{min}$. The beam is allowed to debunch for a period of about 60-200 μsec and then captured with the help of a 37 MHz RF (para-phase) system. Since the magnetic field is continuously changing throughout this period, the beam is captured as quickly as possible with a considerably large RF bucket. This led to substantial beam filamentation in the RF bucket leading to longitudinal emittance dilution. Additional issue in the Booster is the observed jitter in $B_{min}$ relative to the beam injection time which is of the order of 50 μsec. This jitter mainly arises from ComEd power-line frequency fluctuation. Further, the beam capture and acceleration found to partly overlap during this part of the cycle as shown in Fig. 1. A combination of all these effects led to undesirable decreased in beam capture efficiency. Over the years many improvements have been implemented to make the capture more efficient. Yet, the best capture efficiency observed so far is <95% with a longitudinal emittance dilution ≈50%.

Around 2000, the Fermilab long range accelerator program planning started focusing on increasing the beam power on targets for neutrino beams. A staged improvement approach is undertaken to inject beam on all of the Booster cycles and increase the BT per cycle [3, 4]. In this context, the Booster is found to play a significant role in the near future of Fermilab. In 2007, I started investigating possible advantages of beam injection on the deceleration part of the magnet ramp in the Booster and thereby, increase the beam brightness at extraction.

Tests of beam injection on the deceleration part of the magnet ramp in the Booster had been attempted in the past [5]. However, the beam transmission efficiencies turned out to be were very poor and the root causes of the problem were not understood at that time.

This paper proposes a fully developed *Early Beam Injection scheme* which has many advantages over the scheme in current operation. The general principle of the method, beam dynamics simulations and results of the proof of principle experiments are presented. Multi-particle beam dynamics simulations applied to the Booster injection convincingly validates the concepts and the proposed scheme's feasibility.

## PRINCIPLE OF THE EARLY INJECTION SCHEME AND SIMULATIONS

Schematic views of the newly proposed early injection scheme (EIS) along with the currently used scheme (CIS) are shown in Fig. 1. In EIS, the beam is injected at about 150 μsec prior to $\dot{B} = 0$. Following the completion of the injection, the Booster RF system is turned on at a *matched* frequency. Debunching of the beam prior to the start of beam capture is eliminated. The RF capture voltage is increased by changing the para-phase angle from $180^0$ to $0^0$ in ≈ 260 μsec imposing $\dot{P} = 0$ to guarantee iso-adiabatic beam capture in stationary RF buckets with synchrotron oscillation period varying in the range of 125 μsec to 40 μsec. In an ideal case, one demands much longer capture time. Since, the magnetic field is continuously changing the capture time cannot be increased much further. Differential relationship between magnetic field $B$ and the radius $R$ of the orbit is given by, $\Delta R = (R/\gamma_T^2)(\Delta B/B)$ where, $\gamma_T$=5.47, $\Delta B/B = 3.74E-4$ and $R = 75.41$ m for the Booster. Then the radial displacement of the beam due to change in magnetic field is ≈0.9 mm, which is << the diameter of the RF cavity beam pipe (57.2 mm). However, the corresponding change in the RF frequency is ≈13.7 kHz. This RF frequency variation should be taken in to account during beam capture.

We have demonstrated the feasibility of the EIS in the Booster that includes i) beam capture with no emittance growths and no beam losses, ii) beam acceleration from injection energy to the extraction energy and iii) bunch rotation. The 2D- particle tracking simulation code ESME

---

* Work supported by Fermi Research Alliance, LLC under Contract No. De-AC02-07CH11359 with the United States Department of Energy
# cbhat@fnal.gov

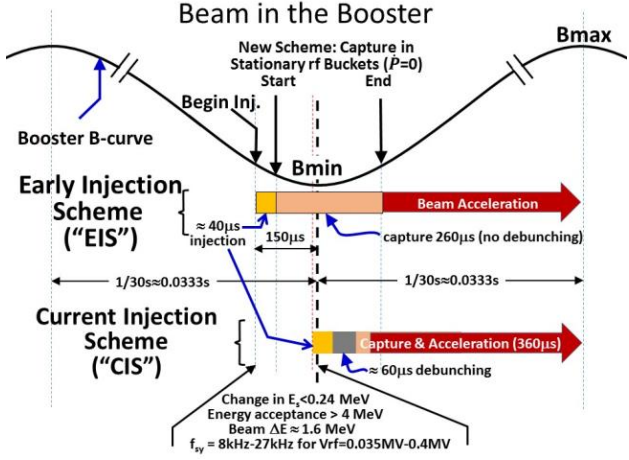

Figure 1: A schematic view of EIS and CIS.

Table 1: Booster parameters used in the simulations

| Parameters | |
|---|---|
| Booster circumference ($2\pi R$) [m] | 473.8 |
| Injection KE [MeV] | 400 |
| Extraction KE [MeV] | 8000 |
| Cycle Time[sec] | 1/15 |
| Beam injection w.r.t. $\dot{B}=0$ [μsec] | 0, -90, -144 |
| Harmonic Number | 84 |
| Transition Gamma $\gamma_T$ | 5.478 |
| $\Delta E$ at Injection [MeV] | 0.8-1.6 [7] |
| Longitudinal Emittance [eV sec] | 0.04 |
| Beam Structure at Injection | 201MHz |
| Number of BT | 1-17 |
| Bunch Intensity [protons/bunch] | 2E10-12E10 |
| Beam transverse radius [cm] | 1.2* |
| Beam pipe (RF) radius [cm] | 2.86* |

*Used in simulations with space charge effects

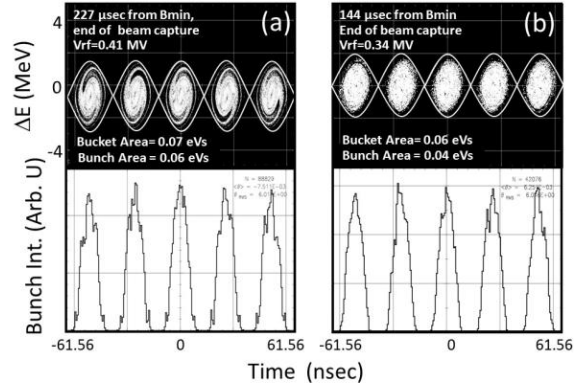

Figure 2: Simulated phase space distributions (top) and time projection (bottom) just before beam acceleration for a) CIS and b) EIS.

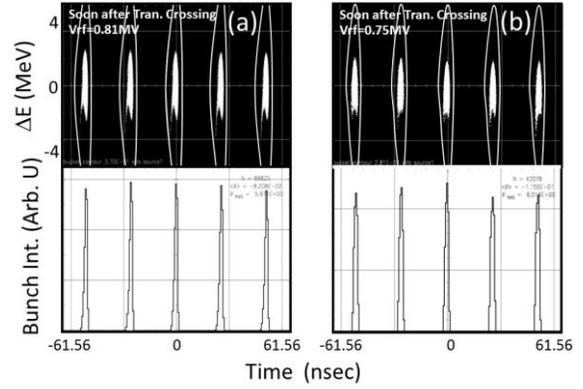

Figure 3: Simulation results soon after transition crossing. Descriptions are similar to that for Fig. 1.

[6] was employed to validate the scheme and to establish the rf manipulation steps needed in the beam experiments. Table 1 shows the machine and beam parameters used in the simulations.

Figure 2 compares the results from the simulated phase space distributions and their time projections after completion of beam capture for 17 BT beam in the EIS and the CIS with and without space-charge effects, respectively. We use measured initial beam energy spread [7] in our simulations. The predicted bunch area, bucket area and the required RF voltage at these instances of capture processes are also shown. It is evident that there is no emittance growth in EIS. On the other hand, in the CIS we have observed ≈50% emittance dilution along with ≈2% beam particles not being captured in the RF bucket which will get lost early in the acceleration cycle.

As we cross the transition energy in the Booster, we make a RF phase jump from ϕ to π-ϕ. Figure 3 compares the predicted snap shot for the particle distributions at about 0.6 msec after the transition crossing. In both cases we observe bucket mis-match leading to a large emittance growths and filamentation. Simulations suggested that this emittance dilution can be minimized by adding a small RF phase kick of about -6 deg after the transition crossing, and hence, we apply this feature only to the EIS cases. This sort of phase displacement is operationally achievable with some minor modifications to the existing Booster LLRF [8]. Once the beam energy is close to the extraction energy we perform snap bunch rotation, i.e., at about 2 msec before the end of the cycle, the RF voltage is increased slowly to ≈ 650 kV to increase the energy spread of the bunches and dropped down rapidly to ≈ 130 kV. This gives minimum energy spread for the beam for slip stacking in the downstream accelerators. The end results for the EIS case after the bunch rotation are shown in Fig. 4. We find that $\Delta E_{RMS}$ can be smaller than that obtained in the current operation by about 30%.

Since there is large longitudinal emittance dilution in the CIS during the beam capture, one expects considerably larger RF bucket area throughout the acceleration cycle to minimize the beam losses, which results in large RF voltage. On the other hand, the required RF power for the EIS is expected to be ≈ 30% smaller than that for the CIS.

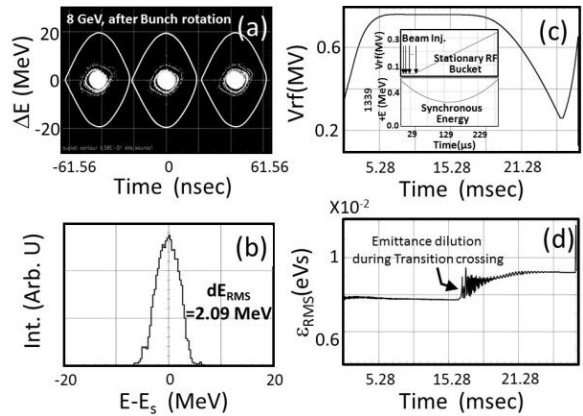

Figure 4: Simulated a) distribution after bunch rotation, b) corresponding energy projection, c) RF voltage curve and d) RMS longitudinal emittance for EIS. The insets in (c) show changing beam synchronous energy and RF curve for the first 300 µsec in the cycle.

## EXPERIMENTS

Proof of principle experiments have been carried out following the guidelines from the simulations on the EIS. During these tests we have scanned the region from 0 µsec to -530 µsec relative to the $B_{min}$. A typical data sample from a wall current monitor, cavity RF voltage, $\dot{B}$ and the beam (for 13 BT) for the first 1 msec in the cycle for beam injection corresponding to -144 µsec is illustrated in Fig. 5(a) (a step at about 600 µsec in the beam signal is due to removal of two bunches in the Booster ring). By setting the RF frequency matched to the injection energy the beam survived through the cycle as shown in Fig. 5(b).

At the time of these experiments the required hardware that allows a good matching between the beam energy and the rf frequency from injection to the $B_{min}$ (the RF frequency should follow $B$ with $\Delta f$ = 13.7 kHz) was in planning stage. Consequently, we intentionally delayed the start of the beam capture at $B_{min}$ as indicated in Fig. 5(a) and the beam capture could not be quite adiabatic. In spite of these issues, the beam acceleration efficiency found to be about 94%. At this time the needed hardwares i) at injection that gives better control on the RF frequency, ii) that can produce additional rf phase kick to achieve better match between beam distribution to the bucket after transition and iii) bunch rotation, are under development. Issues related to transverse dynamics is yet to be addressed.

In conclusion, we have proposed a new injection scheme for the Fermilab Booster which will give lower longitudinal emittance and no beam loss from injection to the extraction energy. We demonstrated the scheme by simulations and with a proof of principle experiment. We need additional LLRF development for full implementation of EIS in operation. Since there is more room (or can be added) for the beam injection in EIS, one can potentially increase the beam intensity in the Booster by injecting more number of Booster turns if the LINAC permits.

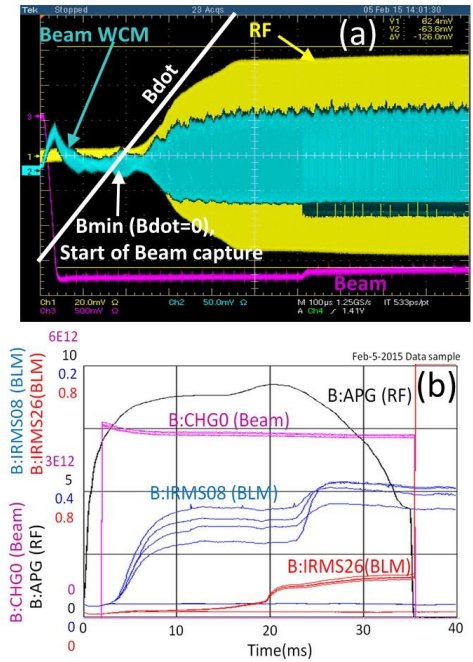

Figure 5: A measurement data on early injection scheme: a) Scope data for the first 1 ms after beam injection (pink trace is -1×Beam) b) 13BT beam, total RF voltage and loss monitor data from injection to extraction without notch in the beam. The beam efficiency was nearly 94% in this case.

I would like to thank W. Pellico, K. Triplett, S. Chaurize, B. Hendrick, and T. Sullivan for their help in the beam studies. My sincere thanks are due to B. Chase for useful discussions during the early stages of this work.